\documentclass[pre,twocolumn,superscriptaddress]{revtex4}

\usepackage{anysize}
\usepackage{graphicx}
\usepackage{epsfig}
\usepackage{verbatim}
\usepackage{eurosym}
\usepackage{amssymb}
\usepackage{amsfonts,amsmath}
\usepackage{bm}
\usepackage{eufrak}
\usepackage[colorlinks]{hyperref}
\usepackage{fancyhdr}
\usepackage{relsize}
\usepackage[T1]{fontenc}
\usepackage{xr}
\usepackage{anysize}
\usepackage{relsize}
\usepackage{mathrsfs}
 \usepackage{newpxtext,newpxmath}
 \usepackage{booktabs,siunitx}
\usepackage{array,booktabs,ragged2e}

\marginsize{1.5cm}{1.5cm}{1.cm}{1.cm}
\hypersetup{colorlinks=true,
	linkcolor=black,
	anchorcolor=black,
	citecolor=black,
	urlcolor=black
}

\providecommand{\be}{\begin{equation}}
\providecommand{\ee}{\end{equation}}
\providecommand{\bea}{\begin{eqnarray}}
\providecommand{\eea}{\end{eqnarray}}
\providecommand{\beas}{\begin{eqnarray*}}
\providecommand{\eeas}{\end{eqnarray*}}

\providecommand{\beni}{\begin{equation*}}
\providecommand{\eeni}{\end{equation*}}

\providecommand{\bw}{\begin{widetext}}
\providecommand{\ew}{\end{widetext}}

\providecommand{\bal}{\begin{aligned}}
\providecommand{\eal}{\end{aligned}}

\def\br{{\bm r}}
\def\bs{{\bm s}}
\def \bsd{\dot{\bs}}
\def \brd{\dot{\br}}

\def\bX{{\bm X}}
\def\bx{{\bm x}}
\def\th{\vartheta}
\def\bell{{\bm \ell}}
\def\la{\lambda}

\makeatletter
\newcommand{\vast}{\bBigg@{2}}
\newcommand{\Vast}{\bBigg@{4}}
\makeatother

\newcommand{\nn}{\nonumber}

\begin{document}

\title{Symmetry reduction of the three-body problem based on Euler angles}
\begin{abstract}
We consider the classical three-body problem with an arbitrary pair potential  which depends on the inter-body distance. A general three-body configuration is  set by three `radial' and three angular variables, which determine the shape and orientation, respectively, of a triangle with the three bodies located at the vertices. The radial variables are given by the distances between a reference body and the other two, and by the angle at the reference body  between the other two. Such radial variables set the  potential energy of the system, and they are reminiscent of the inter-body distance in the two-body problem. On the other hand, the angular variables are  the Euler angles relative to a rigid rotation of the triangle, and they  are analogous to the polar and azimuthal angle of the vector between the two bodies in the two-body problem. We show that the rotational symmetry allows us to obtain a closed set of eight Hamilton  equations of motion, whose generalized coordinates are the thee radial variables and one additional angle, for which we provide the following geometrical interpretation.   
Given a reference body, we consider the plane through it which is orthogonal to the line between the reference and a second body. We show that the angular variable above is the angle between the plane projection of the angular-momentum vector, and the projection of the radius between the reference and the third body.  
\end{abstract}

\author{Michele Castellana}
\affiliation{Laboratoire Physico-Chimie Curie, Institut Curie, CNRS UMR168, 75005 Paris, France}
\pacs{45.50.Pk,45.50.Jf,45.20.Jj}

\maketitle

\setlength{\parskip}{5pt plus 0pt minus 0pt}

\section{Introduction}\label{intro}

The motion of three bodies under their mutual interaction arises in a variety of contexts in physics. The characterization and prediction of this motion, named the three-body problem, has played a fundamental role since the original work of Newton on the Sun-Earth-Moon system \cite{newton1687principia}.
Such a prototypical case of the gravitational three-body problem attracted growing interest in the eighteenth and nineteenth century \cite{gutzwiller1998moon}, both as a  testing ground for the law of universal gravitation, and for its fundamental technological applications. Indeed, the prediction of the relatively quick motion of the Moon on the background sky allowed for reckoning Greenwich mean time from the observed Moon's position with the method of lunar distance, and thus for determining the longitude of the observer  \cite{bowditch2002the,maskelyne1750maskelyne,maskelyne1761journal}. 

Given that an explicit solution proved to be out of reach, multiple studies focused on how  to simplify the  problem by reducing its number of degrees of freedom, in an effort to obtain reliable computations of the ephemerides \cite{poincare1892methodes,szebehely1972gravitational}. { In what follows, we will review some of these reductions in chronological order.} 

Among the most significant symmetry reductions of the problem is the one proposed by Lagrange \cite{lagrange1772essai}. By algebraically manipulating Newton's equations of motion, Lagrange obtained a closed set of four second-order and three first-order differential equations for seven dynamical variables, and the existence of four independent integrals allowed for reducing these equations to a smaller set equivalent to seven, first-order equations. 
Later on, Jacobi presented a reduction to a set of differential equations of equivalent complexity: The problem was first reduced to a set of equations describing the movement of two bodies, which were further simplified by means of the elimination of their orbital nodes \cite{jacobi1843sur}. 

Further studies performed the reduction to a system of equations  in the Hamiltonian form: Radau derived a system of equations for the  distances of two bodies from the third, their angular distances from the line of nodes, and the respective conjugate momenta, which may be effectively reduced to seven first-order equations by using the conservation of the energy  \cite{radau1868sur}. A reduction to a Hamiltonian system with the same number of degrees of freedom was performed by Poincar\'e in terms of Keplerian variables, in an effort to analyze the perturbative series of the reduced Hamiltonian \cite{poincare1892methodes}. A few years later,  Bennett proposed an additional Hamiltonian reduction with four generalized coordinates given by a subset of the cartesian coordinates of two bodies in a reference frame centered in the third  \cite{bennett1905on}. 
A further symmetry reduction was proposed by Van Kampen and Wintner, who derived a set of Hamilton equations for four degrees of freedom, the inter-body distances and an angular variable symmetrical with respect to permutations of the  bodies, by leveraging a set of invariant relations for canonical coordinates \cite{vankampen1937on}. 
{ An additional symmetry reduction suited to perturbative approaches has been proposed in a recent analysis \cite{malige2002partial}: Unlike other studies where the degrees of freedom are reduced by two units at once, this reduction is performed in two separate steps based on the conservation of the direction and norm of the angular-momentum vector, respectively. Although the first step allows for a reduction by one degree of freedom only, it presents some advantages, e.g., it preserves the rotational symmetry, which is broken in other approaches.}
 
Finally,  recent analyses used a combination of differential geometry and geometric-invariant theory to  reduce the equations of motion, and obtained a set of nine first-order differential equations with two constants of motion, whose complexity is thus equivalent to the reduced equations above \cite{hsiang2007kinematic,sydnes2013geometric}. 

In this analysis, we consider the  three-body problem with an arbitrary pair potential which depends on the inter-body distance, and propose a symmetry-reduction  procedure based on Euler angles, inspired by an analogy with the two-body problem.  A three-body configuration is described by a triangle, whose vertices correspond to the three bodies,  and a general triangle conformation is described by three `radial' and three angular variables. 
First, the `radial' variables determine the triangle shape, and they are given by the length of two sides of the triangle and the angle between them---in the gravitational three-body problem the latter angle corresponds to the astronomical elongation, i.e., the angle at one body  between the other two  \cite{bowditch2002the}.
Such radial variables set the potential energy of the system, and they are therefore reminiscent of the modulus of the vector between the two bodies in the two-body problem with a radial potential. Second, the triangle orientation is set by three angular variables---the Euler angles which describe a rigid rotation of the triangle, which are analogous to the polar and azimuthal angle of the vector above in the two-body problem. 

By using the Hamiltonian formalism \cite{poincare1892methodes}, we show that the rotational-symmetry differential conditions on the Hamiltonian identify directly a set of   variables in terms of which we can write eight reduced Hamilton equations: These  variables are given by the radial coordinates, one additional angular variable, and the respective conjugate momenta. Combined with the conservation of the total energy, our analysis provides a system of seven first-order equations, whose complexity is equivalent to that of the studies previously discussed.  Finally, we study the geometric interpretation of the additional angular variable above: Given the plane through a reference body orthogonal to the line between the reference and a second body, we show that such variable is the angle between the plane projection of the line between the reference and the third body, and the plane projection of angular-momentum vector. 

The rest of the paper is organized as follows. In Section \ref{sec1} we revisit the two-body problem with the framework of Hamiltonian mechanics, and illustrate how  rotational symmetry  allows us to obtain two first-order equations of motion for the inter-body radius and its conjugate momentum. In Section \ref{sec2} we show how to apply this analysis to the three-body problem: In Section \ref{sec2a} we introduce a set of radial and angular variables reminiscent of those in the two-body example and derive the corresponding Hamiltonian formalism, whose rotational symmetry is analyzed in Section \ref{sec2b}. In Section \ref{sec2c} we derive the resulting reduced equations of motion and Hamiltonian, whose features and geometric interpretation are discussed in Section \ref{dis}. 

\section{The two-body problem revisited}\label{sec1}

In order to illustrate our symmetry-reduction procedure with an example, consider two bodies with masses $m_1$, $m_2$ and coordinates $\bX_1$, $\bX_2$  in an inertial reference frame, which interact through a  potential $U(|\bX_1 - \bX_2|)$. Setting 
\be\label{eq4}
\br \equiv \bX_1 - \bX_2,
\ee
 the equation of motion for $\br$ reads 
\be\label{eq61}
\mu\,  \ddot{\br} = - \frac{\partial U(r)}{\partial \br},
\ee
where  $\dot{\, }$ denotes the time derivative,
\be\label{eq69}
\mu \equiv \frac{m_1 m_2}{m_1+m_2} 
\ee
is the reduced mass, and 
\be\label{eq90}
r \equiv \left| \br \right|.
\ee
Equation (\ref{eq61}) is the Euler-Lagrange equation of the Lagrangian
\be
L(\br , \dot{\br}) = \frac{\mu}{2} \dot{\br}^2 - U(r).
\ee
We now write  $\br$ in terms of radial and angular variables,  setting
\be\label{eq95}
\br   =  R_z(\gamma) R_x(\beta)  \br_0,
\ee
where $\br_0 \equiv (0,0,r)$, and the matrices 
\bea\label{eq62}
R_z(\theta) &\equiv& \left(\begin{array}{ccc}
\cos\theta & -\sin\theta & 0\\
\sin \theta & \cos\theta & 0\\
0 & 0 & 1
\end{array}\right),\\
\label{eq63}
R_x(\theta) &\equiv& \left(\begin{array}{ccc}
1 & 0 & 0\\
0 & \cos\theta & -\sin\theta\\
0 & \sin\theta & \cos\theta
\end{array}\right)
\eea
represent a rotation by  $\theta$ about the  $z$ and $x$ axes, respectively. In Eq. (\ref{eq95}) we have denoted the polar and azimuthal angle by $\beta$ and $\gamma$, respectively, in order to draw an analogy with the three-body problem in Section \ref{sec2}. We adopt ${\bm q}\equiv (r, \beta, \gamma)$ as generalized coordinates, in terms of which re-write the Lagrangian formulation above. The resulting Hamiltonian 
\be\label{eq65}
H({\bm q}, {\bm p}) = \frac{p_r^2}{2 \mu	} + \frac{p_\beta^2 + \csc^2 \beta \, p_\gamma^2}{2 \mu \, r^2} + U(r)
\ee
depends on the angular variables $\beta$, $p_\beta$ and $p_\gamma$  through the square modulus of the angular momentum ${\bm \ell}^2 = p_\beta^2 + \csc^2 \beta \, p_\gamma^2$,  where ${\bm \ell} \equiv \br \times \mu \dot{\br} = ( \cos \gamma \, p_\beta - \cot \beta  \sin \gamma \, p_\gamma  ,  \sin \gamma \, p_\beta+ \cot \beta  \cos \gamma \, p_\gamma, p_\gamma )$. 
Given that ${\bm \ell}^2$ is a constant of motion, the Hamilton equations for $r$ and $p_r$,  are decoupled from the equations for the angular variables, and the problem is thus reduced to a single degree of freedom.  

 The observation above that $H$ depends on the angular variables and conjugate momenta through ${\bm \ell}^2$ can be easily made by direct inspection of Eq. (\ref{eq65}). However, for systems with a more complex Hamiltonian structure such as the three-body problem, the identification of the dependence of $H$ on its angular variables by direct inspection is not straightforward, see Section \ref{sec2}. For this reason, we will show that such dependence can be worked out with an alternative procedure, i.e., by leveraging the rotational invariance of $H$ \cite{poincare1892methodes}: in what follows, we will detail this procedure for the two-body problem as an illustrative example. The rotational invariance of $H$ implies \cite{goldstein2004classical}
\be\label{eq12}
\left[ H, \bm \ell \right] = 0,
\ee
where 
\be\label{eq13}
 \left[ \,\, , \,  \right] \equiv \sum_i \left( \frac{\partial}{\partial q_i} \frac{\partial}{\partial p_i} - \frac{\partial}{\partial p_i} \frac{\partial}{\partial q_i} \right)
\ee
is the Poisson bracket, and the summation runs over the generalized coordinates and conjugate momenta ${\bm p} \equiv (p_r, p_\beta, p_\gamma)$.   Equation  (\ref{eq12}) implies the existence of three constants of motion---the three components of the angular momentum $\bm \ell$. Given that these constants are not in involution with each other, the condition (\ref{eq12}) does not allow us to lower by three the number of degrees of freedom of the system \cite{poincare1892methodes}. However, it is possible to find two combinations of the angular-momentum components, e.g., $\ell_z$ and $\bell^2$, which are both constants of motion and in involution with each other:
\bea\label{eq67}
\left[ H, \ell_z \right] &=& 0,\\ \label{eq68}
\left[ H, \bell^2 \right]&=& 0.
\eea

The differential conditions (\ref{eq67}) and (\ref{eq68}) will allow us to reduce the number of degrees of freedom of the problem by two units: The condition (\ref{eq67}) implies
\be\label{eq40}
\frac{\partial H}{\partial \gamma} = 0. 
\ee
On the other hand, Eq. (\ref{eq68}) yields the following partial differential equation
\be
p_\beta  \frac{\partial H}{\partial \beta}+p_\gamma^2 \cot \beta  \csc ^2\beta \,  \frac{\partial H}{\partial p_\beta}=0.  
\ee
which can be solved directly, yielding 
\be\label{eq66}
H({\bm q}, {\bm p}) = {\mathscr H}(p_\beta^2 + \csc^2 \beta \, p_\gamma^2, r, p_r, p_\gamma). 
\ee
Equation (\ref{eq66}) reproduces the result (\ref{eq65}), i.e., that the Hamiltonian depends on $\beta$, $p_\beta$  through the constant of motion ${\bm \ell}^2$, which allows us to write an equation of motion for a single degree of freedom as discussed above.

\section{Three-body problem}\label{sec2}

In this Section we will perform the symmetry reduction of the three-body problem, proceeding along the lines of the illustrative example above.

\subsection{Hamiltonian formalism for radial variables and Euler angles}\label{sec2a}

Consider three bodies with masses $m_1$, $m_2$, $m_3$ and coordinates $\bX_1$, $\bX_2$, $\bX_3$ in an intertial reference frame, where bodies $i$ and $j$ interact through a pair potential $U_{ij}(|\bX_i - \bX_j|)$. The equations of motion read
\be\label{eq1}
m_i \ddot{\bX}_i = - \sum_{j\neq i} \frac{\partial U_{ij}(|\bX_i - \bX_j|)}{\partial \bX_j},
\ee
for $i=1,2,3$, and $U_{ij}(r) = U_{ji}(r)$ is an arbitrary pair potential between bodies $i$ and $j$, which depends on their positions  through the inter-body distance $|\bX_i - \bX_j|$. { We now introduce a new set of variables $\br$ and $\bs$, where $\br$ is given by Eq. (\ref{eq4}),  and
\be\label{eq4b}
\bs \equiv \bX_3- \bX_2.
\ee
In the gravitational three-body problem for the Earth-Moon-Sun system, the variables $\br$ and $\bs$ coincide with the heliocentric cartesian coordinates,  where bodies 1, 2 and 3 are the Earth, Sun and Moon, respectively \cite{poincare1952oeuvres,morbidelli2002modern}. }

We consider Eq. (\ref{eq1}) for $i=2$ and multiply it by $m_1$, consider Eq. (\ref{eq1}) for $i=1$ and multiply it by $m_2$, subtract the two equations, and proceed along the same lines for $i=2$ and $3$. We obtain the following equations of motion for $\bm r$ and $\bm s$
\be\label{eq2}
\bal
\mu\, \ddot{\br} &= - \frac{\partial U_{12}(r)}{\partial \br} - \frac{\mu}{m_2} \frac{\partial U_{23}(s)}{\partial \bs}  - \frac{\mu}{m_1} \frac{\partial U_{13}(\left|\br - \bs\right|)}{\partial\br },\\
\nu\,  \ddot{\bs} &= - \frac{\partial U_{23}(s)}{\partial \bs} - \frac{\nu}{m_2} \frac{\partial U_{12}(r)}{\partial \br} - \frac{\nu}{m_3} \frac{\partial U_{13}(\left|\br - \bs\right|)}{\partial \bs},
\eal
\ee
where $M \equiv m_1 + m_2 + m_3$ is the total mass, the reduced masses $\mu$ and $\nu$ are defined by Eq. (\ref{eq69}) and by
\be
\nu \equiv \frac{m_2 m_3}{m_2+m_3},
\ee
$r$ is given by Eq. (\ref{eq90}), and $s \equiv \left| \bs \right|$. 

Equations (\ref{eq2}) can be regarded as the equations of motion of two interacting bodies with coordinates $\br$ and $\bs$ and masses $\mu$ and $\nu$, respectively. 
In order to carry out their symmetry reduction, we will study Eq. (\ref{eq2}) with the Lagrangian formalism. In this regard,  we set
\be\label{eq17}
V(r, s, \left| \br - \bs \right|) \equiv U_{12}(r) + U_{13}(\left| \br - \bs \right|)+ U_{23}(s), 
\ee
and consider the Lagrangian 
\be\label{eq6} 
\bal
L(\br, \bs, \brd, \bsd) \equiv&  \frac{1}{2 M} \big[ m_1 (m_2+m_3) \brd^2 + m_3(m_1+m_2) \bsd^2 \\
&- 2 m_1 m_3 \brd \cdot \bsd \big] - V(r, s, \left| \br - \bs \right|)
\eal
\ee
{ which, in the gravitational three-body problem, is known to yield the Hamiltonian and canonical equations of motion in heliocentric coordinates \cite{poincare1952oeuvres}.} From the expression (\ref{eq6}), it is straightforward to show that the Euler-Lagrange equations of  $L$ imply Eqs. (\ref{eq2}): as a result, we may take $L$ as the Lagrangian of the two-body system.

\begin{figure*}
\centering\includegraphics[scale=1.5]{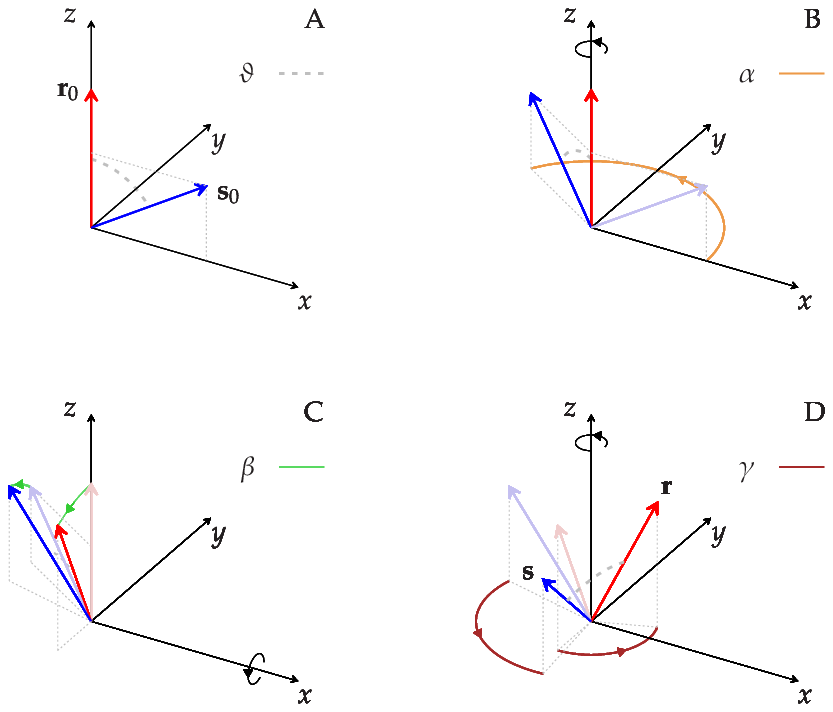}
\caption{
Geometrical constriction based on Euler angles. (A): The vectors $\br_0 = (0,0,r)$ (in red) and $\bs_0 = s(\sin \th, 0, \cos \th)$ (blue) denote a reference configuration  for the vectors $\br = \bX_1 - \bX_2$ and $\bs = \bX_3 - \bX_2$, respectively, which determine the relative positions of the three bodies. A general configuration of $\br$ (red) and $\bs$ (blue) is obtained by applying a sequence of rigid rotations to the triangle formed by  $\br_0$ and $\bs_0$: A rotation by an angle $\alpha$ about the $z$ axis  (panel B, orange arc), a rotation by $\beta$ around the $x$ axis (panel C, green arcs), and a rotation by $\gamma$ about the $z$ axis (panel D, brown arcs). The opaque arrows in each panel denote the configuration in the preceding panel. 
\label{fig1}}
\end{figure*}

We will now  leverage the rotational symmetry of the problem by making the change of coordinates depicted in Fig. \ref{fig1}. Given $\br_0 \equiv (0,0,r)$, $\bs_0 \equiv s (\sin \th , 0 , \cos \th)$, we write a general configuration $\br$, $\bs$ as a rigid rotation of the triangle formed by $\br_0$ and $\bs_0$: 
\bea\label{eq7}
\br  & = &  R_z(\gamma) R_x(\beta) R_z(\alpha) \br_0, \\ \label{eq8}
\bs & = &  R_z(\gamma) R_x(\beta) R_z(\alpha) \bs_0, 
\eea
where $R_z(\theta)$ and $R_x(\theta)$ represent a rotation by  $\theta$ about the $z$ and $x$ axes, and they are given by Eqs. (\ref{eq62}) and (\ref{eq63}), respectively. 
 
Equations (\ref{eq7}) and (\ref{eq8}) allow us to write the six variables $\br$, $\bs$ in terms of  $r$, $s$, $\th$, $\alpha$, $\beta$ and $\gamma$. The variables $r$, $s$ and $\th$ determine the shape of the triangle formed by $\br$ and $\bs$, whose vertices coincide with the three bodies. On the other hand, the Euler angles $\alpha$, $\beta$ and $\gamma$ characterize a rigid rotation of the triangle \cite{goldstein2004classical}, and thus set the triangle's orientation. 
An analogous set of variables occurs in the symmetry reduction of the two-body problem above, where the potential depends on the modulus of the radius $r$  only, not on the angular variables $\beta$ and $\gamma$ which determine  the direction of $\br$. Along the same lines, the rotational symmetry for the three-body problem implies that the potential $V(r, s, \left| \br - \bs \right|) = V(r, s, \sqrt{r^2+s^2 - 2 \, r s \cos \th})$ is a function of the variables $r$, $s$ and $\th$ only, which  will thus be denoted by `radial' variables, as opposed to the angular variables $\alpha$, $\beta$, $\gamma$. 

We set 
${\bm q} = (r, s, \th, \alpha, \beta, \gamma)$, and rewrite the Lagrangian $L$ as a function of $\bm q$ and $\dot{\bm q}$. By deriving both sides of Eqs. (\ref{eq7}) and (\ref{eq8}) with respect to $t$, we obtain $\dot{\br}$ and $\dot{\bs}$ as functions of ${\bm q}$ and $\dot{\bm q}$
\be\label{eq16}
\bal
\dot{\br} = \dot{\br}({\bm q}, \dot{\bm q}), \,\,
\dot{\bs} = \dot{\bs}({\bm q}, \dot{\bm q}), 
\eal
\ee
and substitute Eq. (\ref{eq16}) in Eq. (\ref{eq6}). 
We then construct the Hamiltonian formalism for $L$ by   introducing the conjugate momenta 
\be\label{eq15}
{\bm p} = \frac{\partial L}{\partial \dot{\bm q}},
\ee
and obtain 
\bw
\bea\label{eq9}\nn
H({\bm q}, {\bm p}) & \equiv &  \sum_i p_i \dot{q}_i - L  \\ \nn 
& =&  \frac{1}{2}\vast\{ \frac{p_r^2}{\mu} + \frac{p_s^2}{\nu} + \vast(  \frac{1}{\mu \,  r^2} + \frac{1}{\nu s^2}  - \frac{2 \cos \th}{m_2 r s} \vast) p_\th^2 \\ \nn
&&+  \vast[  \frac{\cot^2 \beta}{\mu \,  r^2} - 2\frac{\cot \th \csc \th}{m_2 r s} + \vast(\frac{1}{\nu s^2} + \frac{1}{\mu \,  r^2}\vast) \csc^2 \th - \frac{1}{\mu \,  r^2} + 2 \sin \alpha \cot \beta \vast( -\frac{\cot \th}{\mu \,  r^2}  + \frac{\csc \th}{m_2 r s}\vast) \vast] p_\alpha^2 \\ \nn
&&+ \frac{1}{\mu \,  r^2} p_\beta^2 + \frac{\csc^2 \beta}{\mu \,  r^2} p_\gamma^2 \vast\} +  \frac{\cos \th }{m_2 } p_r p_s - \frac{\sin \th}{m_2}\left( \frac{p_r}{s}+\frac{p_s}{r}\right) p_\th    \\ \nn 
& &  -\frac{\cos \alpha \cot \beta \sin \th}{m_2 r} p_s p_\alpha - \frac{\sin \alpha \sin \th}{m_2 r} p_s p_\beta + \frac{\cos \alpha \sin \th \csc \beta}{m_2 r} p_s p_\gamma  + \frac{\cos \alpha \cot \beta}{r^2 s}\vast( \frac{s}{\mu} - \frac{r \cos \th}{m_2}\vast) p_\th p_\alpha \\ \nn
&& + \frac{\sin \alpha }{r}\vast( \frac{1}{\mu \,  r} - \frac{\cos \th}{s m_2}\vast) p_\th p_\beta  + \frac{\cos \alpha \csc \beta}{r}\vast( -\frac{1}{\mu \,  r} 
+  \frac{\cos \th}{s m_2}\vast) p_\th p_\gamma  
+ \frac{\cos \alpha \csc \th}{r} \vast( -\frac{1}{m_2 s} + \frac{\cos \th}{\mu \,  r} \vast) p_\alpha p_\beta \\ 
&& + \frac{\csc \beta}{r} \vast[ - \frac{\cot \beta}{\mu \,  r} + \sin \alpha \csc \th \vast( -\frac{1}{m_2 s} + \frac{\cos \th}{\mu\,  r}\vast) \vast] p_\alpha p_\gamma  + V(r, s, \sqrt{r^2+s^2 - 2 \, r s \cos \th}),
\eea
\ew
which is the Hamiltonian for the radial coordinates, the Euler angles, and the respective conjugate momenta. 

{
Finally, we observe that Eqs. (\ref{eq15}) and (\ref{eq9}) may alternatively be obtained by considering the generalized coordinates $Q = (\br, \bs)$, constructing their conjugate momenta $P = ( \frac{\partial L}{\partial \brd}, \frac{\partial L}{\partial \bsd})$ and Hamiltonian, and  introducing a canonical transformation $(Q,P) \rightarrow (q,p)$ with generating function of the second type $F  = F_2(Q,p)-\sum_i q_i p_i$ \cite{goldstein2004classical}, where $F_2(Q,p) = \sum_i f_i(Q) P_i$  and ${\bm q} = {\bm f}(Q)$ is the transformation described by Eqs. (\ref{eq7}) and (\ref{eq8}). 
} 

\subsection{Rotational-symmetry conditions}\label{sec2b}

Proceeding along the lines of Section \ref{sec1}, the rotational symmetry of the Hamiltonian (\ref{eq9}) will be written in terms of Poisson brackets and  angular momentum. 

Denoting the coordinate of the center of mass (CM) by ${\bm X}_{\rm CM} \equiv (m_1 \bX_1 + m_2 \bX_2 + m_3 \bX_3)/M$ and the body coordinates in the CM reference frame by 
\be\label{eq10}
\bx_i \equiv \bX_i - \bX_{\rm CM},
\ee 
we express the three variables $\bx_i$ in terms of $\br$ and $\bs$ by means of Eqs. (\ref{eq4}), (\ref{eq4b}), (\ref{eq10}) and  the condition $m_1 \bx_1 + m_2 \bx_2 + m_3 \bx_3 = 0$. We  obtain
\be\label{eq5}
\bal
\bx_1 &= \frac{(m_2+m_3) \br - m_3 \bs}{M} ,\\
\bx_2 &= -\frac{m_1 \br + m_3 \bs}{M} ,\\
\bx_3 &= \frac{ -m_1 \br  + (m_1+m_2) \bs }{M}. 
\eal
\ee

As a result, the angular momentum in the CM reference frame can be written as
\bw
\bea
\label{eq20}
\bell & \equiv &  \sum_{i=1}^3 \bx_i  \times m_i \dot{\bx}_i \\ \nn
& =& \frac{1}{M} \big\{ \br \times \big[m_1(m_2+m_3) \dot{\br} - m_1 m_3 \dot{\bs} \big] +\bs \times \big[- m_1 m_3 \dot{\br} + m_3(m_1+m_2) \dot{\bs}  \big]\big\}\\ \nn
& =&  ( \csc \beta  \sin \gamma  \, p_\alpha + \cos \gamma \, p_\beta - \cot \beta \sin \gamma \,  p_\gamma,   - \csc \beta  \cos \gamma  \, p_\alpha + \sin \gamma \, p_\beta + \cot \beta \cos \gamma \,  p_\gamma , p_\gamma),
\eea
\ew
where in the second line we used Eqs. (\ref{eq5}), and in the third line we expressed $\dot{\br}$, $\dot{\bs}$ as functions of the conjugate momenta by using Eqs. (\ref{eq7}), (\ref{eq8}), (\ref{eq16}) and (\ref{eq15}). The rotational symmetry of the Lagrangian (\ref{eq6}) implies the differential condition (\ref{eq12}), where the Poisson bracket is given by Eq. (\ref{eq13}), and the summation in Eq. (\ref{eq13}) runs over the generalized coordinates $r$, $s$, $\th$, $\alpha$, $\beta$, $\gamma$ and their conjugate momenta. 

Proceeding along the lines of Section \ref{sec1}, we consider the quantities $ \ell_z$ and ${\bm \ell}^2$, which are both in involution and and constants of motion, see Eqs. (\ref{eq67}) and (\ref{eq68}).  While Eq. (\ref{eq67})  implies that the Hamiltonian is independent of $\gamma$, i.e., Eq. (\ref{eq40}), the condition  (\ref{eq68}) results in a more involved partial differential equation for $H$. In order to write explicitly this equation, we express $\bell^2$ in terms of the conjugate momenta by using Eq. (\ref{eq20}),
\be\label{eq21}
\bell^2 = \csc^2\beta\,  (p_\alpha^2 + p_\gamma^2) - 2 \cos \beta \csc^2 \beta \, p_\alpha p_\gamma + p_\beta^2,
\ee
we substitute Eq. (\ref{eq21}) in Eq. (\ref{eq68}), and  obtain
\bw 
\bea\label{eq22}\nn
(p_\alpha - \cos \beta\, p_\gamma)\csc^2 \beta \frac{\partial H}{\partial \alpha} + p_\beta \frac{\partial H}{\partial \beta} + (p_\alpha - \cos \beta \, p_\gamma )(\cos \beta \, p_\alpha - p_\gamma)\csc^3 \beta \frac{\partial H}{\partial p_\beta}&=& 0.  
\eea
\ew

We  solve the linear, first-order partial differential equation  (\ref{eq22})  in the variables $\alpha$, $\beta$ and $p_\beta$ with the method of characteristics \cite{courant1966methods}. The characteristic curves $\alpha(\la)$, $\beta(\la)$ and $p_\beta(\la)$ are defined in terms of the parameter $\la$, and  they satisfy 
\bw
\bea \label{eq23a}
\cfrac{d\alpha(\lambda) }{d\la} & = &  \left[p_\alpha -  p_\gamma \cos \beta(\lambda)  \right]\csc^2 \beta(\lambda),\\\label{eq23b}
\cfrac{d\beta(\lambda) }{d\la} & = &  p_\beta(\lambda),\\ \label{eq23c}
\cfrac{dp_\beta(\lambda)}{d\la} & = &  \left[p_\alpha -  p_\gamma \cos \beta(\lambda)  \right]   \left[p_\alpha \cos \beta(\lambda) -  p_\gamma   \right]  \csc^3 \beta(\lambda). 
\eea
\ew
In order to solve the system of ordinary differential equations above, we eliminate the parameter $\la$ by combining Eqs. (\ref{eq23b}) and (\ref{eq23c}): 
\be\label{eq24}
\frac{d p_\beta}{d \beta} = \frac{ (p_\alpha -  p_\gamma \cos \beta  )(p_\alpha \cos \beta -  p_\gamma )  \csc^3 \beta}{p_\beta}, 
\ee
where the characteristic curves are now parametrized in terms of $\beta$ rate than in terms of $\lambda$. By integrating Eq. (\ref{eq24}), we obtain
\be\label{eq25}
p_\beta = \varsigma \sqrt{ -  (p_\alpha^2 + p_\gamma^2 - 2 \, p_\alpha p_\gamma \cos \beta) \csc \beta^2 + C_1}, 
\ee
where $\varsigma=\pm 1$  denotes the sign of $p_\beta$ throughout the rest of our analysis, and $C_1$ is an integration constant which will be determined in what follows. 
In order to determine $\beta(\lambda)$, we substitute Eq. (\ref{eq25}) in Eq. (\ref{eq23b}), and obtain
\be\label{eq26}
\frac{d \beta(\lambda)}{d \lambda} = \varsigma \sqrt{ -  [p_\alpha^2 + p_\gamma^2 - 2 \, p_\alpha p_\gamma \cos \beta(\lambda) ] \csc \beta(\lambda) ^2 + C_1}. 
\ee
By restricting our analysis to values of $\lambda$ for which $p_\beta(\lambda)$ does not change sign, Eq. (\ref{eq26}) implies
\be
\bal
\varsigma \lambda& = \int \frac{d \beta}{\sqrt{ -  (p_\alpha^2 + p_\gamma^2 - 2 \, p_\alpha p_\gamma \cos \beta ) \csc \beta ^2 + C_1}}&\\
&=-\frac{1}{\sqrt{C_1}} \arcsin \frac{C_1 \cos \beta - p_\alpha p_\gamma }{\sqrt{(C_1-p_\alpha^2)(C_1-p_\gamma^2)}} + C_2,  &
\eal
\ee
which can be solved for $\beta(\lambda)$,  yielding
\bw
\be\label{eq27}
\beta(\lambda )  =  \arccos\vast\{ \frac{1}{C_1} \vast[ p_\alpha p_\gamma - \varsigma \sqrt{(C_1-p_\alpha^2)(C_1-p_\gamma^2)}  \sin\left( \sigma(\lambda) \right)  \vast] \vast\},
\ee
\ew
where we have set 
\be
 \sigma(\lambda) \equiv \sqrt{C_1} (\lambda - C_2). 
\ee
By substituting Eq. (\ref{eq27}) in Eq. (\ref{eq25}), we obtain $p_\beta(\lambda)$: 
\bw
\be\label{eq28}
p_\beta(\lambda) =   \frac{\varsigma \, \sqrt{C_1 (C_1-p_\alpha^2)(C_1-p_\gamma^2)}\cos \left( \sigma(\lambda) \right)}{\sqrt{C_1^2- \left[p_\alpha p_\gamma - \varsigma \sqrt{(C_1-p_\alpha^2)(C_1-p_\gamma^2)}\sin\left( \sigma(\lambda) \right) \right] ^2}}. 
\ee
\ew
Finally,  we substitute Eq. (\ref{eq27}) in Eq. (\ref{eq23a}) and obtain $\alpha(\lambda)$: 
\bw
\be\label{eq37}
\bal
\alpha(\lambda)  &= C_1 \int d \lambda \,  \frac{p_\alpha C_1 - p_\gamma \left[ p_\alpha p_\gamma - \varsigma\sqrt{(C_1-p_\alpha^2)(C_1-p_\gamma^2)} \sin\left( \sigma(\lambda) \right) \right]}{C_1^2- \left[p_\alpha p_\gamma - \varsigma \sqrt{(C_1-p_\alpha^2)(C_1-p_\gamma^2)}\sin\left( \sigma(\lambda) \right) \right] ^2} & \\
& =  \arctan \frac{2 \sqrt{C_1} \cos^2\left( \frac{\sigma(\lambda) }{2}\right)\left[ p_\alpha (C_1 - p_\gamma^2) \tan\left( \frac{\sigma(\lambda) }{2}\right)  + \varsigma \, p_\gamma \sqrt{(C_1-p_\alpha^2)(C_1-p_\gamma^2)} \right] }{p_\gamma^2 (p_\alpha^2 - C_1) - C_1 (p_\alpha^2-C_1) \cos\left( \sigma(\lambda)\right) - \varsigma \, p_\alpha p_\gamma \sqrt{(C_1-p_\alpha^2)(C_1-p_\gamma^2)} \sin\left( \sigma(\lambda)\right) } + C_3. &
\eal 
\ee
\ew
The differential equations (\ref{eq23a})-(\ref{eq23c}) imply that the Hamiltonian is constant across the characteristic curves \cite{courant1966methods}, in particular
\be\label{eq29}
H(\alpha(\lambda), \beta(\lambda), p_\beta(\lambda)) = H(\alpha(C_2), \beta(C_2), p_\beta(C_2)),
\ee
where we indicated the dependence of the Hamiltonian on the variables $\alpha$, $\beta$ and $p_\beta$ only to simplify the notation. To obtain the  explicit dependence of the left-hand side of Eq. (\ref{eq29}) on $\alpha(\lambda)$, $\beta(\lambda)$, $p_\beta(\lambda)$, we express $\alpha(C_2)$, $\beta(C_2)$, $p_\beta(C_2)$ in the right-hand side (RHS) as functions of $\alpha(\lambda)$, $\beta(\lambda)$, $p_\beta(\lambda)$ by using the solution above for the characteristic curves. To achieve this, we combine Eqs. (\ref{eq27}) and  (\ref{eq28}) and solve them for $\sin(\sigma(\lambda))$ and $C_1$, which are thus expressed as functions of $\beta(\lambda)$ and $p_\beta(\lambda)$. 
Setting
\be\label{eq42}
\Phi(\beta, p_\beta, p_\alpha, p_\gamma) \equiv 
\csc^2\beta\,  (p_\alpha^2 + p_\gamma^2) - 2 \cos \beta \csc^2 \beta \, p_\alpha p_\gamma+ p_\beta^2,
\ee
the solution for $C_1$ and $\sin (\sigma(\lambda))$ reads
\bea\label{eq30}
C_1 &=& \Phi(\beta(\lambda), p_\beta(\lambda), p_\alpha, p_\gamma),\\ \label{eq38}
\sin(\sigma(\lambda)) & = & \varsigma \frac{p_\alpha p_\gamma - C_1 \cos( \beta(\lambda)) }{\sqrt{(C_1-p_\alpha^2)(C_1-p_\gamma^2)}},
\eea
where Eqs. (\ref{eq21}), (\ref{eq42}) and (\ref{eq30}) show that the integration constant $C_1$ coincides with  the square modulus of the  angular momentum on the characteristic curve. Combining Eqs. (\ref{eq27}) and (\ref{eq28}) for $\lambda = C_2$ with Eq. (\ref{eq30}), we obtain that that $\beta(C_2)$, $p_\beta(C_2)$ depend on $\alpha(\lambda)$, $\beta(\lambda)$, $p_\beta(\lambda)$  through the combination $\Phi(\beta(\lambda), p_\beta(\lambda), p_\alpha, p_\gamma)$. 

Finally, we carry out the same analysis for $\alpha(C_2)$: we consider Eq. (\ref{eq37}) for $\lambda = C_2$, and obtain 
\be\label{eq91}
\alpha(C_2)  =  \varsigma \arctan \frac{2 \sqrt{C_1}   p_\gamma \sqrt{(C_1-p_\alpha^2)(C_1-p_\gamma^2)}}{p_\gamma^2 (p_\alpha^2 - C_1) - C_1 (p_\alpha^2-C_1) }  + C_3.
\ee
The first term on the RHS of Eq. (\ref{eq91}) depends on $\beta(\lambda)$ and $p_\beta(\lambda)$  through the combination $\Phi$. As far as the second term $C_3$ is concerned, we substitute Eq. (\ref{eq30}) in Eq. (\ref{eq37}) and express all trigonometric functions in terms of $\sin(\sigma(\lambda))$, which we rewrite according to Eq. (\ref{eq38}). As a result, Eq. (\ref{eq37}) allows us to write $C_3$ as a function of $\alpha(\lambda)$, $\beta(\lambda)$ and $p_\beta(\lambda)$. Both terms in the RHS of Eq. (\ref{eq91}) are thus expressed in terms of $\alpha(\lambda)$, $\beta(\lambda)$, $p_\beta(\lambda)$, and after a few algebraic manipulations we obtain:
\bw
\be\label{eq43}
\alpha(C_2) =\\  
\Psi(\alpha(\lambda), \beta(\lambda), p_\beta(\lambda), p_\alpha, p_\gamma)   + \varsigma \arctan \frac{p_\gamma \sqrt{\Phi(\beta(\lambda), p_\beta(\lambda), p_\alpha, p_\gamma) - p_\alpha^2}}{\sqrt{\Phi(\beta(\lambda), p_\beta(\lambda), p_\alpha, p_\gamma) [\Phi(\beta(\lambda), p_\beta(\lambda), p_\alpha, p_\gamma) - p_\gamma^2]}},
\ee
\ew
where
\be\label{eq73}
\Psi(\alpha, \beta, p_\beta, p_\alpha, p_\gamma) \equiv \alpha - \arctan \frac{p_\gamma - \cos \beta \, p_\alpha }{\sin \beta \, p_\beta}.
\ee

By combining Eq. (\ref{eq29})  with Eqs. (\ref{eq27}) and (\ref{eq28}) evaluated at $\la = C_2$ and with Eq. (\ref{eq43}), we obtain
\bw
\be \label{eq72}
\bal
H(\alpha(\lambda), \beta(\lambda), p_\beta(\lambda)) &=&\\ 
 H\Bigg(\Psi(\alpha(\lambda), \beta(\lambda), p_\beta(\lambda), p_\alpha, p_\gamma) + \varsigma \arctan \frac{p_\gamma \sqrt{\Phi(\beta(\lambda), p_\beta(\lambda), p_\alpha, p_\gamma) - p_\alpha^2}}{\sqrt{\Phi(\beta(\lambda), p_\beta(\lambda), p_\alpha, p_\gamma) [\Phi(\beta(\lambda), p_\beta(\lambda), p_\alpha, p_\gamma) - p_\gamma^2]}},&&\\   
 \arccos\frac{p_\alpha p_\gamma}{\Phi(\beta(\lambda), p_\beta(\lambda), p_\alpha, p_\gamma)}, &&\\
 \varsigma \frac{\sqrt{\Phi(\beta(\lambda), p_\beta(\lambda), p_\alpha, p_\gamma) [\Phi(\beta(\lambda), p_\beta(\lambda), p_\alpha, p_\gamma)-p_\alpha^2][\Phi(\beta(\lambda), p_\beta(\lambda), p_\alpha, p_\gamma)-p_\gamma^2]}}{\sqrt{[\Phi(\beta(\lambda), p_\beta(\lambda), p_\alpha, p_\gamma)]^2-p_\alpha^2 p_\gamma^2}}
 \Bigg).&&
\eal
\ee
\ew
By writing explicitly the dependence on $p_\alpha$ and $p_\gamma$, Eq. (\ref{eq72}) yields
\bw
\be\label{eq41}
\bal
H(\alpha, \beta, p_\beta, p_\alpha, p_\gamma) &=&\\
 H\Bigg(\Psi(\alpha, \beta, p_\beta, p_\alpha, p_\gamma) + \varsigma \arctan \frac{p_\gamma \sqrt{\Phi(\beta, p_\beta, p_\alpha, p_\gamma) - p_\alpha^2}}{\sqrt{\Phi(\beta, p_\beta, p_\alpha, p_\gamma) [\Phi(\beta, p_\beta, p_\alpha, p_\gamma) - p_\gamma^2]}}, \arccos\frac{p_\alpha p_\gamma}{\Phi(\beta, p_\beta, p_\alpha, p_\gamma)}, &&\\
\varsigma \frac{ \sqrt{\Phi(\beta, p_\beta, p_\alpha, p_\gamma)[\Phi(\beta, p_\beta, p_\alpha, p_\gamma)-p_\alpha^2][\Phi(\beta, p_\beta, p_\alpha, p_\gamma)-p_\gamma^2]}}{\sqrt{[\Phi(\beta, p_\beta, p_\alpha, p_\gamma)]^2-p_\alpha^2 p_\gamma^2}}\Bigg)&\equiv&\\
 \mathscr{H}(\Phi(\beta, p_\beta, p_\alpha, p_\gamma), \Psi(\alpha, \beta, p_\beta, p_\alpha, p_\gamma), p_\alpha, p_\gamma),&&
\eal
\ee
\ew
where  in the third line we rewrote the overall variable dependence of the second line in terms of a function $\mathscr H$ of $\Phi(\beta, p_\beta, p_\alpha, p_\gamma)$, $\Psi(\alpha, \beta, p_\beta, p_\alpha, p_\gamma)$, $p_\alpha$ and $p_\gamma$. 

Equation (\ref{eq41}) constitutes the solution of the partial differential equation (\ref{eq22}), and it is the analog of Eq. (\ref{eq66}) for the two-body problem. Although in both the two- and three-body problem  $\mathscr H$ is a function of the total angular momentum, Eq. (\ref{eq41}) shows that  in the three-body problem $\mathscr H$ depends on an additional angular variable $\Psi$, whose interpretation will be discussed below.  

\subsection{Reduced equations of motion}\label{sec2c}

The differential conditions (\ref{eq40}) and (\ref{eq41}) will allow us to reduce the number of degrees of freedom by two units, i.e., to write a closed set of  equations of motion which involve only four variables as opposed to the six degrees of freedom of the Lagrangian (\ref{eq6}).  

To achieve this, we will follow the general procedure discussed in \cite{poincare1892methodes}. We invert Eqs. (\ref{eq42}) and (\ref{eq73}), i.e., we express $\beta$ and $p_\beta$ as functions of $\Phi, \Psi, \alpha, p_\alpha$ and  $p_\gamma$: we introduce the functions $\beta(\varphi,\psi,\alpha, p_\alpha, p_\gamma)$ and $p_\beta(\varphi,\psi,\alpha, p_\alpha, p_\gamma)$, defined by 
\be\label{eq51}
\bal
\Phi(\beta(\varphi,\psi,\alpha, p_\alpha, p_\gamma), p_\beta( \varphi, \psi, \alpha, p_\alpha, p_\gamma), p_\alpha, p_\gamma) & =  \varphi,&  \\
\Psi(\alpha, \beta(\varphi,\psi,\alpha, p_\alpha, p_\gamma), p_\beta( \varphi, \psi,\alpha, p_\alpha, p_\gamma), p_\alpha, p_\gamma) & =  \psi,&  
\eal
\ee
where $\varphi$ and $\psi$ are two independent variables and, given that Eqs. (\ref{eq42}) and (\ref{eq73}) may have multiple solutions for $\beta$ and $p_\beta$, Eq. (\ref{eq51}) should be regarded as a local inversion. 

The equation of motion for $\Psi$ reads
\bw
\be\label{eq50}
\bal
\dot{ \Psi} &= \frac{\partial \Psi}{\partial \alpha}\frac{\partial H}{\partial p_\alpha} +  \frac{\partial \Psi}{\partial \beta}\frac{\partial H}{\partial p_\beta} -  \frac{\partial \Psi}{\partial p_\beta}\frac{\partial H}{\partial\beta} - \frac{\partial \Psi}{\partial p_\alpha} \frac{\partial H }{\partial \alpha}&\\
& = \frac{\partial H}{\partial p_\alpha} +  \frac{\partial \Psi}{\partial \beta}\frac{\partial H}{\partial p_\beta} -  \frac{\partial \Psi}{\partial p_\beta}\frac{\partial H}{\partial\beta} +\frac{\partial \Psi}{\partial p_\alpha} \vast[   \frac{p_\beta\sin^2 \beta}{p_\alpha - \cos \beta\, p_\gamma} \frac{\partial H}{\partial \beta} + (\cos \beta \, p_\alpha - p_\gamma)\csc \beta \frac{\partial H}{\partial p_\beta}\vast]&\\
& = \frac{\partial H}{\partial p_\alpha} + \vast[ - \frac{\partial \Psi}{\partial p_\beta}+ \frac{p_\beta\sin^2 \beta}{p_\alpha - \cos \beta\, p_\gamma} \frac{\partial \Psi}{\partial p_\alpha}\vast]\frac{\partial H}{\partial \beta}+\vast[ \frac{\partial \Psi }{\partial \beta} + (\cos \beta \, p_\alpha - p_\gamma)\csc \beta  \frac{\partial \Psi }{\partial p_\alpha}   \vast] \frac{\partial H}{\partial p_\beta} ,& 
\eal
\ee
\ew
where in the first line we used Hamilton equations of motion and Eq. (\ref{eq40}), and in the second line Eqs.  (\ref{eq22}) and  (\ref{eq73}).  
We rewrite the terms in brackets in the last line of Eq. (\ref{eq50}) as functions of the derivatives of $\mathscr H$ by proceeding as follows: We derive Eqs. (\ref{eq51}) with respect to $p_\alpha$, set 
\be
\bal\label{eq76}
\varphi&=\Phi(\beta, p_\beta, p_\alpha, p_\gamma),&\\
 \psi&= \Psi(\alpha, \beta, p_\beta, p_\alpha, p_\gamma),&
\eal
\ee
 and obtain
\be\label{eq52}
\left\{
\begin{array}{ccc}
\displaystyle
\frac{\partial \Phi}{\partial \beta} \frac{\partial \beta}{\partial p_\alpha} + \frac{\partial \Phi}{\partial p_\beta} \frac{\partial p_\beta}{\partial p_\alpha}  + \frac{\partial \Phi}{\partial p_\alpha} & = & 0 ,\vspace{2mm}\\
\displaystyle
\frac{\partial \Psi}{\partial \beta} \frac{\partial \beta}{\partial p_\alpha} + \frac{\partial \Psi}{\partial p_\beta} \frac{\partial p_\beta}{\partial p_\alpha}  + \frac{\partial \Psi}{\partial p_\alpha} & = & 0,
\end{array}
\right.
\ee
which we solve for $\partial \beta /\partial p_\alpha$ and $\partial p_\beta /\partial p_\alpha$. 
By using Eqs. (\ref{eq42}) and (\ref{eq73}), we observe that  
\be
\bal\label{eq93}
 \frac{\partial \beta}{\partial p_\alpha} & =   - \frac{\partial \Psi}{\partial p_\beta}+ \frac{p_\beta\sin^2 \beta}{p_\alpha - \cos \beta\, p_\gamma} \frac{\partial \Psi}{\partial p_\alpha} ,&\\
 \frac{\partial p_\beta}{\partial p_\alpha} & =   \frac{\partial \Psi }{\partial \beta} + (\cos \beta \, p_\alpha - p_\gamma)\csc \beta  \frac{\partial \Psi }{\partial p_\alpha},&
\eal
\ee
solves Eq. (\ref{eq52}). Combined with Eq. (\ref{eq50}), Eq. (\ref{eq93}) implies  
\be\label{eq55}
\dot{\Psi}  =  \frac{\partial H}{\partial \beta}\frac{\partial \beta}{\partial p_\alpha}  +\frac{\partial H}{\partial p_\beta}\frac{\partial p_\beta}{\partial p_\alpha} + \frac{\partial H}{\partial p_\alpha}. 
\ee

We will now write the RHS of Eq. (\ref{eq55}) in terms of the reduced Hamiltonian $\mathscr H$. To achieve this,  we use the definition (\ref{eq51}) and rewrite Eq. (\ref{eq41}) as
\be \label{eq56}
\bal
H(\alpha, \beta(\varphi,\psi, \alpha,p_\alpha, p_\gamma), p_\beta(\varphi,\psi, \alpha,p_\alpha, p_\gamma),p_\alpha,p_\gamma) &=\\
\mathscr{H}(\varphi,\psi, p_\alpha, p_\gamma). &
\eal
\ee

By  deriving Eq. (\ref{eq56}) with respect to $p_\alpha$ and fixing $\varphi$ and $\psi$ according to Eqs. (\ref{eq76}), we reconstruct the RHS of Eq. (\ref{eq55}):
\bea\nn
 \frac{\partial \mathscr H}{\partial p_\alpha}  & =  &\frac{\partial H}{\partial \beta}\frac{\partial \beta}{\partial p_\alpha}  +\frac{\partial H}{\partial p_\beta}\frac{\partial p_\beta}{\partial p_\alpha} + \frac{\partial H}{\partial p_\alpha}  \\\label{eq58}
 &=& \dot{\Psi}. 
\eea
Equation (\ref{eq58}) is the equation of motion for $\Psi$, and it has the form of a Hamilton equation of motion with Hamiltonian $\mathscr H$, where $p_\alpha$ is the conjugate momentum of $\Psi$. 
\\

Proceeding along the same lines, we work out the equation of motion for $p_\alpha$: We have
\be\label{eq57}
\dot{p}_\alpha = - \frac{\partial H}{\partial \alpha}, 
\ee
and express the RHS of Eq. (\ref{eq57}) as a function of $\mathscr H$. We derive both sides of Eq. (\ref{eq56}) with respect to $\psi$, set $\varphi$ and $\psi$ according to Eqs. (\ref{eq76}), and obtain
\be\label{eq77}
\frac{\partial \mathscr H}{\partial \Psi} = \frac{\partial H}{\partial \beta}\frac{\partial \beta}{\partial \Psi}  +\frac{\partial H}{\partial p_\beta}\frac{\partial p_\beta}{\partial \Psi}.
\ee
Proceeding along the lines of Eq. (\ref{eq52}), we work out $\partial \beta/\partial \Psi$ and $\partial p_\beta/\partial \Psi$ in Eq. (\ref{eq77}): We derive Eqs. (\ref{eq51}) with respect to $\psi$, impose Eqs. (\ref{eq76}),  solve the resulting linear system 
\be\label{}
\left\{
\begin{array}{ccc}
\displaystyle
\frac{\partial \Phi}{\partial \beta} \frac{\partial \beta}{\partial \Psi} + \frac{\partial \Phi}{\partial p_\beta} \frac{\partial p_\beta}{\partial \Psi}  & = & 0 ,\vspace{2mm}\\
\displaystyle
\frac{\partial \Psi}{\partial \beta} \frac{\partial \beta}{\partial \Psi} + \frac{\partial \Psi}{\partial p_\beta} \frac{\partial p_\beta}{\partial \Psi}   & = & 1, 
\end{array}
\right.
\ee
and obtain
\be\label{eq80}
\bal
 \frac{\partial \beta}{\partial \Psi} & =   \frac{p_\beta \sin^2\beta}{p_\gamma \cos \beta - p_\alpha},&\\
 \frac{\partial p_\beta}{\partial \Psi} & =  \frac{p_\gamma - p_\alpha \cos \beta}{\sin \beta}.&
\eal
\ee

By substituting  Eq. (\ref{eq80}) in Eq. (\ref{eq77}), we reconstruct the derivative of $H$ with respect to $\alpha$:  
\be\label{eq81}
\bal
\frac{\partial \mathscr H}{\partial \Psi} & =  \frac{\partial H}{\partial \beta}\frac{p_\beta \sin^2\beta}{p_\gamma \cos \beta - p_\alpha}  +\frac{\partial H}{\partial p_\beta}\frac{p_\gamma - p_\alpha \cos \beta}{\sin \beta}&\\
& =  \frac{\partial H}{\partial \alpha},&
\eal
\ee
where in the second line we used Eq. (\ref{eq22}). By substituting  Eq. (\ref{eq81}) in  Eq. (\ref{eq57}), we obtain the equation of motion for $p_\alpha$: 
\be\label{eq82}
\dot{p}_\alpha = - \frac{\partial \mathscr H}{\partial \Psi}.
\ee
\vspace{1mm}\\
Put together, Eqs. (\ref{eq58}) and (\ref{eq82}) have the form of the Hamilton equations for two the generalized coordinate $\Psi$ and its conjugate momentum $p_\alpha$.

The equations of motion for the remaining  variables $r$, $s$, $\th$ and their conjugate momenta can be written directly in terms of $\mathscr H$ by combining Hamilton equations of motion with Eq. (\ref{eq41}): 
\be\label{eq94}
\dot{q} = \frac{\partial \mathscr H}{\partial p}, \, \, \dot{p} = -\frac{\partial \mathscr H}{\partial q}, 
\ee
with $q = r,s,\th$ and $p= p_r, p_s, p_\th$, respectively. \vspace{5mm}\\

Finally, we work out the  expression for the reduced Hamiltonian $\mathscr H$. Because both sides of Eq. (\ref{eq56}) are independent of $\alpha$, $\mathscr H$ can be obtained from $H$ by setting $\alpha$ to an arbitrary value in Eq. (\ref{eq56}). We choose  $\alpha = \psi$, given that for this value of $\alpha$ the expression (\ref{eq73}) simplifies, thus the inverse functions $\beta$ and $p_\beta$ in Eq. (\ref{eq51}) can be worked out easily:
\be\label{eq85}
\bal
\beta(\varphi, \psi, \psi, p_\alpha, p_\gamma) &= \arccos\frac{ p_\gamma}{p_\alpha},&\\
p_\beta(\varphi, \psi, \psi, p_\alpha, p_\gamma) &= \varsigma \sqrt{\varphi-p_\alpha^2}.&
\eal
\ee
We substitute Eq. (\ref{eq85}) in Eq. (\ref{eq56}), set $\varphi = \Phi$ and $\psi = \Psi$, and obtain the reduced Hamiltonian
\bw
\bea\nn
\mathscr{H} & = & \frac{1}{2 \mu} \left(p_r^2 + \frac{\Phi}{r^2} \right) + \frac{p_s^2}{2 \nu} + \left( \frac{1}{\mu\, r^2}+\frac{1}{\nu s^2} - \frac{2 \cos \th}{m_2 r s } \right) \frac{p_\th^2}{2} + \csc^2 \th \left( \frac{1}{\nu s^2} + \frac{\cos( 2 \th)}{\mu\, r^2} - \frac{2 \cos \th}{m_2 r s}\right) \frac{p_\alpha^2}{2} - \frac{\sin \th}{m_2 r} p_s p_\th\\ \nn
&&  + \frac{p_r}{m_2}\left(p_s \cos \th - \frac{p_\th \sin \th}{s} \right)  + \varsigma \frac{\sqrt{\Phi-p_\alpha^2}}{m_2 \,  \mu \, r^2 s}[p_\alpha (m_2 s \cot \th - \mu\, r \csc \th)\cos \Psi + (m_2 s\, p_\th  \\ \label{eq86}
&& -\mu\, r \cos \th \, p_\th - \mu\, r s \sin \th\, p_s ) \sin \Psi]+ U_{12}(r) + U_{13}(\sqrt{r^2  + s^2 - 2 \, r s \cos \th}) + U_{23}(s).
\eea
\ew

To summarize, the reduced equations of motion  can be written explicitly by combining Eqs. (\ref{eq58}), (\ref{eq82}) and (\ref{eq94}), and they read
\be\label{eq71}
\left\{
\begin{array}{llll}
\displaystyle \dot{r} = \frac{\partial \mathscr H}{\partial p_r}, & \displaystyle \dot{s} = \frac{\partial \mathscr H}{\partial p_s},& \displaystyle \dot{\th} = \frac{\partial \mathscr H}{\partial p_\th}, & \displaystyle \dot{\Psi} = \frac{\partial \mathscr H}{\partial p_\alpha}, \vspace{2mm}\\
\displaystyle
\dot{p}_r = -\frac{\partial \mathscr H}{\partial r},  & \displaystyle \dot{p}_s = -\frac{\partial \mathscr H}{\partial s},  & \displaystyle \dot{p}_\th = -\frac{\partial \mathscr H}{\partial \th}, &\displaystyle \dot{p}_\alpha = -\frac{\partial \mathscr H}{\partial \Psi}, 
\end{array}
\right.
\ee
where the reduced Hamiltonian is given by Eq. (\ref{eq86}), and  $\varsigma=\pm 1$ for $p_\beta \gtrless 0$, respectively. Equations (\ref{eq86}) and (\ref{eq71}) constitute the main result of this paper. 

{
\vspace{1cm}
While Eqs. (\ref{eq71}) determine the time dependence of the  radial variables $r, s, \th$ and conjugate momenta, the remaining angular variables $\alpha$, $\beta$ and $\gamma$ can be determined from an additional set of differential and algebraic equations---see Appendix \ref{app1} for details. Indeed, the RHSs of Hamilton equations of motion of $\alpha$ and $\gamma$
\be \label{eq112}
\dot{\alpha}  =  \frac{\partial H}{\partial p_\alpha}, \, \, \dot{\gamma}  =  \frac{\partial H}{\partial p_\gamma},
\ee
can be written as follows:
\bw
\bea\label{eq113}
\bal
\cfrac{\partial H}{\partial p_\alpha} & = &\\
 \frac{\partial \mathscr H}{\partial p_\alpha}+\frac{2 [(p_\alpha^2-\Phi ) \sin ^2(\alpha -\Psi )-p_\alpha^2] [-p_\alpha^3+p_\alpha (p_\alpha^2-\Phi ) \sin ^2(\alpha -\Psi )+p_\alpha p_\gamma^2-p_\gamma \tau  \Lambda]}{p_\alpha (p_\alpha^3-p_\alpha p_\gamma^2+2 p_\gamma \tau  \Lambda)-(p_\alpha^2+p_\gamma^2) (p_\alpha^2-\Phi ) \sin ^2(\alpha -\Psi )} \frac{\partial \mathscr H}{\partial \Phi}
&&\\
- \frac{  \varsigma \cos (\alpha -\Psi ) (\tau  \Lambda-p_\alpha p_\gamma) \sqrt{(\Phi -p_\alpha^2) \{[(\Phi -p_\alpha^2) \sin ^2(\alpha -\Psi )+p_\alpha^2]^2-(p_\alpha p_\gamma-\tau  \Lambda)^2\}}}{[p_\gamma (\Phi -p_\alpha^2) \sin ^2(\alpha -\Psi )+p_\alpha \tau  \Lambda]^2+(\Phi -p_\alpha^2) \cos ^2(\alpha -\Psi ) \{[(\Phi -p_\alpha^2) \sin ^2(\alpha -\Psi )+p_\alpha^2]^2-(p_\alpha p_\gamma-\tau  \Lambda)^2\}} \frac{\partial \mathscr H}{\partial \Psi},&&\\
\eal
\eea
\bea\label{eq114}
\bal
\cfrac{\partial H}{\partial p_\gamma} & = & \\
\cfrac{\partial \mathscr H}{\partial p_\gamma}   -\frac{ \{(p_\alpha^2-\Phi ) \cos [2 (\alpha - \Psi )]+p_\alpha^2+\Phi \} [p_\gamma (p_\alpha^2-\Phi ) \cos [2 (\alpha - \Psi) ]-p_\alpha^2 p_\gamma+2 p_\alpha \tau  \Lambda+p_\gamma \Phi ]}{2 (p_\alpha^2+p_\gamma^2) (p_\alpha^2-\Phi ) \sin ^2(\alpha -\Psi )-2 p_\alpha (p_\alpha^3-p_\alpha p_\gamma^2+2 p_\gamma \tau  \Lambda)} \frac{\partial \mathscr H}{\partial \Phi} &&\\
- \frac{  \varsigma \cos (\alpha -\Psi ) [(\Phi -p_\alpha^2) \sin ^2(\alpha -\Psi )+p_\alpha^2] \sqrt{(\Phi -p_\alpha^2) \{[(\Phi -p_\alpha^2) \sin ^2(\alpha -\Psi )+p_\alpha^2]^2-(p_\alpha p_\gamma-\tau  \Lambda)^2\}}}{[p_\gamma (\Phi -p_\alpha^2) \sin ^2(\alpha -\Psi )+p_\alpha \tau  \Lambda]^2+(\Phi -p_\alpha^2) \cos ^2(\alpha -\Psi ) \{[(\Phi -p_\alpha^2) \sin ^2(\alpha -\Psi )+p_\alpha^2]^2-(p_\alpha p_\gamma-\tau  \Lambda)^2\}} \frac{\partial \mathscr H}{\partial \Psi},&&
\eal
\eea
\ew
where  
\be\label{eq116}
\tau = \pm1
\ee
 and
\bw
\be\label{eq115}
\Lambda \equiv \sqrt{(\Phi- p_\alpha^2  ) \sin ^2(\alpha -\Psi ) [(\Phi- p_\alpha^2 ) \sin ^2(\alpha -\Psi )+ p_\alpha^2 - p_\gamma^2 ]}.
\ee
\ew
Overall, Eqs. (\ref{eq112}), (\ref{eq113}) and (\ref{eq114}) constitute a closed set of differential equations which depend on the reduced variables $r, s, \th$ and $\Psi$, their conjugate momenta, the constants of motion $\Phi$ and $p_\gamma$, and on the Euler angle $\alpha$. Once the reduced variables and conjugate momenta are obtained from Eqs. (\ref{eq71}),  Eqs. (\ref{eq112}), (\ref{eq113}) and (\ref{eq114}) determine the time dependence of  $\alpha$ and $\gamma$, while the remaining Euler angle $\beta$ is obtained from the algebraic relation
\be\label{eq110}
\cos \beta = \frac{p_\alpha  p_\gamma - \tau \Lambda}{( \Phi-p_\alpha^2 ) \sin ^2(\alpha -\Psi )+ p_\alpha^2}. 
\ee

}

\section{Discussion}\label{dis}

The reduced equations of motion (\ref{eq71}) constitute a set of eight  Hamilton equations for the generalized coordinates $r$, $s$, $\th$, $\Psi$ and their conjugate momenta $p_r$, $p_s$, $p_\th$ and $p_\alpha$, with Hamiltonian $\mathscr H$. Together with the conservation of the energy, $\dot{\mathscr H} = 0$, these equations of motion can be reduced to a system of seven first-order equations, whose complexity is thus tantamount to that of the reductions discussed in Section \ref{intro}.

In what follows, we will discuss some physical and geometrical features of the  reduced system (\ref{eq71}). First, in the limit where the mass and pair interactions of one of the three bodies vanish, $\mathscr H$ reproduces the Hamiltonian of the two-body problem discussed in Section \ref{sec1}. Indeed, it can be shown that, for $m_3 \rightarrow 0$, the conjugate momenta $p_s$, $p_\th$ and $p_\alpha$  defined by Eq. (\ref{eq15}) are $\sim m_3$, thus Eq. (\ref{eq86})  reproduces the two-body-problem Hamiltonian (\ref{eq65}) with $U = U_{12}$. 
 
An additional  feature of the reduced Hamiltonian $\mathscr H$ is that it depends on the sign of $p_\beta$, see Eq. (\ref{eq86}). Given a set of initial conditions for $r$, $s$, $\th$, $\Psi$ and their conjugate momenta, the appropriate sign $\varsigma$ in $\mathscr H$ is chosen according to the sign of $p_\beta$ at the initial time. The reduced equations of motion can  then  be integrated forward in time, keeping the same $\mathscr H$ as long as $p_\beta$ does not change sign. Given that $\mathscr H$ must be a continuous function of time, a change of sign in $p_\beta$ must correspond to the vanishing of  $\Phi - p_\alpha^2$: Indeed, if this were not the case, the second line in Eq. (\ref{eq41}) implies that $\mathscr H$ has a discontinuous jump at the instant of time where $p_\beta$ vanishes. Because the time evolution of $p_\alpha$ is determined by the reduced equations of motion (\ref{eq71}), such equations allow us to determine the instant of time where $\Phi - p_\alpha^2$, and thus $p_\beta$, vanishes, and  to extend the  integration beyond that time by reversing the sign $\varsigma$ in $\mathscr H$. { The sign $\tau$ in Eqs. (\ref{eq113}), (\ref{eq114}) and (\ref{eq110}) may be determined proceeding  along the same lines. }
  
Finally, we will discuss the physical and geometric interpretation of the angular variables $\th$ and $\Psi$ that appear in the reduced equations. A natural physical interpretation for $\th$ is obtained by applying  our analysis  to the Sun-Earth-Moon system \cite{gutzwiller1998moon}, and denoting by body $1$ the Sun, $2$  the Earth, and $3$ the Moon. In this case, the angle $\th$ becomes the lunar elongation from the Sun, which determines the percentage of the Moon surface that is illuminated as seen from Earth, and sets the Moon phases \cite{bowditch2002the}. Given that our reduced equations contain both  the Earth-Moon distance $s$ and the lunar elongation $\th$, they constitute a minimal, reduced system which predicts the size and shape of the illuminated lunar surface as seen from Earth. 

\begin{figure}
\vspace{5mm}
\centering\includegraphics[scale=1.6]{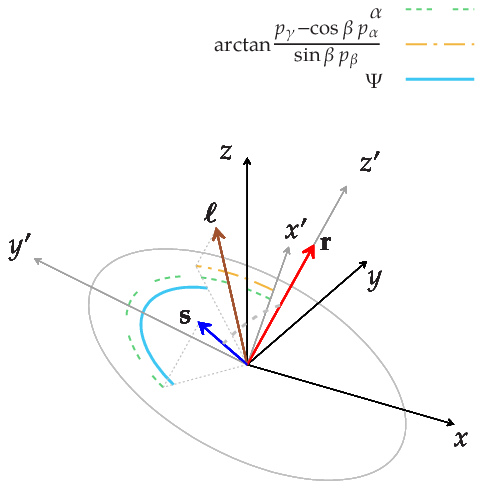}
\caption{
Geometric interpretation of the generalized coordinate $\Psi$ in the reduced equations of motion. Given the $\br$, $\bs$ vectors of Fig. \ref{fig1} (red and blue, respectively),  the axes $x'$, $y'$, $z'$ (gray) are obtained by applying a rotation by $\beta$ about the $x$ axis and a rotation by $\gamma$  about the $z$ axis to $x$, $y$, $z$, respectively. We show the angle $\alpha$ (dashed green curve)  between the projection of $\bs$ on the $x' y'$ plane (gray circle) and $x'$. The projection of the angular-momentum vector $\bm \ell$  (brown) on the  $x' y'$ plane forms an  angle $\arctan[(p_\gamma - \cos \beta \, p_\alpha)/(\sin \beta \, p_\beta)]$  with $x'$ (dashed yellow curve), and  $\Psi$ (solid light-blue curve) is the difference between the two angles above. 
\label{fig2}}
\end{figure}

As far as the variable $\Psi$ is concerned, its geometric interpretation will be clarified in what follows---see Fig. \ref{fig2}. We apply the rotation $R_z(\gamma)R_x(\beta)$ to the axes  $x$, $y$, $z$  of the reference frame in Fig. \ref{fig1}, and obtain the axes $x'$, $y'$ and $z'$. First, the geometric construction in Fig. \ref{fig1} implies that the angle $\alpha$ in Eq. (\ref{eq73}) is the angle between $x'$ and the projection of $\bs$ on the  $x'\, y'$ plane. 
Second, we observe that the argument of $\arctan$ in  Eq. (\ref{eq73}) can be rewritten as
\be
 \frac{p_\gamma - \cos \beta \, p_\alpha }{\sin \beta \, p_\beta} =  \frac{{\bm \ell}_{ y'}}{{\bm \ell}_ { x'}},
\ee
where ${\bm \ell}_{ x'}$ and ${\bm \ell}_{ y'}$ are the projections of $\bm \ell$ on $x'$ and $y'$, respectively. It follows that $\arctan[(p_\gamma - \cos \beta \, p_\alpha)/(\sin \beta \, p_\beta)]$ is the angle between $x'$ and the projection of $\bm \ell$ on the  $x'\,y'$ plane, modulo $\pi$. 
As a result, $\Psi$ is given by the angle between the projection of $\bs$ and the projection of $\bm \ell$ in the   $x'\,y'$ plane, modulo $\pi$. 

\appendix
{
\section{Equations for the Euler angles}\label{app1}

In what follows, we will derive Eqs. (\ref{eq113}), (\ref{eq114}), and the algebraic relation (\ref{eq110}). 

To achieve this, we consider the Hamilton equations of motion (\ref{eq112}) and express their RHSs as functions of $\mathscr H$. By deriving Eq. (\ref{eq56}) with respect to  $\varphi$, $\psi$, $p_\alpha$ and $p_\gamma$, and setting $\varphi$ and $\psi$ according to Eqs. (\ref{eq76}), we obtain
\bea\label{eq100}
\left\{
\begin{array}{lll}
\cfrac{\partial \mathscr H}{\partial \Phi} & = & \cfrac{\partial H}{\partial \beta} \cfrac{\partial \beta}{\partial \Phi} +  \cfrac{\partial H}{\partial p_\beta} \cfrac{\partial p_\beta}{\partial \Phi},\\
\cfrac{\partial \mathscr H}{\partial \Psi} & = & \cfrac{\partial H}{\partial \beta} \cfrac{\partial \beta}{\partial \Psi} +  \cfrac{\partial H}{\partial p_\beta} \cfrac{\partial p_\beta}{\partial \Psi},\\
\cfrac{\partial \mathscr H}{\partial p_\alpha} & = & \cfrac{\partial H}{\partial \beta} \cfrac{\partial \beta}{\partial p_\alpha} +  \cfrac{\partial H}{\partial p_\beta} \cfrac{\partial p_\beta}{\partial p_\alpha}+  \cfrac{\partial H}{\partial p_\alpha},\\
\cfrac{\partial \mathscr H}{\partial p_\gamma} & = & \cfrac{\partial H}{\partial \beta} \cfrac{\partial \beta}{\partial p_\gamma} +  \cfrac{\partial H}{\partial p_\beta} \cfrac{\partial p_\beta}{\partial p_\gamma} +  \cfrac{\partial H}{\partial p_\gamma}.
\end{array}
\right.
\eea
\vspace{9cm}\\

In Eqs. (\ref{eq100}), the derivatives of $\beta$ and $p_\beta$ with respect to $\Phi$, $\Psi$, $p_\alpha$ and $p_\gamma$ may be determined proceeding along the same lines as Section \ref{sec2c}. First, the derivatives with respect to $p_\alpha$ and $\Psi$ are given by Eqs. (\ref{eq93}) and (\ref{eq80}), respectively. Second, the derivatives with respect to $\Phi$ are obtained by deriving Eqs. (\ref{eq51}) with respect to $\varphi$,  imposing Eqs. (\ref{eq76}), and solving the resulting linear system of equations for $\frac{\partial \beta}{\partial\Phi}$ and $\frac{\partial p_\beta}{\partial \Phi}$. The result is
\bw
\bea
\bal\label{eq103}
\frac{\partial \beta }{\partial \Phi} & =  \frac{ (p_\gamma - p_\alpha \cos \beta )\sin^3 \beta }{(p_\alpha - p_\gamma  \cos \beta ) [p_\alpha ^2+\cos (2 \beta ) (p_\alpha^2 - p_\beta^2 ) -4 p_\alpha p_\gamma \cos \beta + p_\beta^2+2 p_\gamma ^2 ]},& \\
\frac{\partial p_\beta }{\partial \Phi} & =  \frac{ p_\beta \sin ^2\beta }{p_\alpha^2+\cos (2 \beta ) (p_\alpha^2-p_\beta^2) -4 p_\alpha p_\gamma \cos \beta +p_\beta ^2+2 p_\gamma^2}&. 
\eal
\eea
\ew
Proceeding along the same lines, we derive Eqs. (\ref{eq51}) with respect to $p_\gamma$, impose Eqs. (\ref{eq76}),  solve the resulting linear systems for $\frac{\partial \beta}{\partial p_\gamma}$ and $\frac{\partial p_\beta}{\partial p_\gamma}$, and obtain

\bea
\bal\label{eq104}
\frac{\partial \beta }{\partial p_\gamma} & =  \frac{\sin \beta }{p_\gamma \cos \beta - p_\alpha},& \\
\frac{\partial p_\beta }{\partial p_\gamma} & =  0.& 
\eal
\eea
\vspace{.4cm}\\

Finally, we use Eqs. (\ref{eq93}), (\ref{eq80}), (\ref{eq103}) and (\ref{eq104}) in Eqs. (\ref{eq100}), solve Eq. (\ref{eq100}) for  
\be
\cfrac{\partial H}{\partial \beta}, \, \cfrac{\partial H}{\partial p_\beta}, \, \cfrac{\partial H}{\partial p_\alpha}, \,\cfrac{\partial H}{\partial p_\gamma},
\ee
and obtain

\bw
\bea\label{eq105}
\frac{\partial H } {\partial p_\alpha}  & = &  \frac{\partial \mathscr H}{\partial p_\alpha} 
+  2   (p_\alpha - p_\gamma \cos \beta ) \csc^2 \beta \frac{\partial \mathscr H}{\partial \Phi} 
+ \frac{p_\beta \sin (2 \beta)}{2[(p_\gamma - p_\alpha \cos \beta)^2+p_\beta^2 \sin^2\beta]}\frac{\partial \mathscr H}{\partial \Psi},\\ \label{eq111}
\frac{\partial H } {\partial p_\gamma}  & = &\frac{\partial \mathscr H}{\partial p_\gamma} 
+  2   (p_\gamma - p_\alpha \cos \beta ) \csc^2 \beta \frac{\partial \mathscr H}{\partial \Phi} 
- \frac{p_\beta \sin\beta }{(p_\gamma - p_\alpha \cos \beta )^2 + p_\beta^2 \sin^2 \beta} \frac{\partial \mathscr H}{\partial \Psi}. 
\eea
\ew

The RHSs of Eqs. (\ref{eq105}) and (\ref{eq111}) depend  only on the reduced variables $r, s, \th, \Psi$ and conjugate momenta,   on the constants of motion $\Phi$ and $p_\gamma$, and on $\beta$, $p_\beta$. In what follows, we will write  $\beta$ and $p_\beta$ in terms of $\Phi$, $\Psi$, $p_\alpha$, $p_\gamma$ and $\alpha$, and thus express (\ref{eq105}) and (\ref{eq111}) in terms of the reduced variables, their conjugate momenta, the constants of motion and $\alpha$ only. To achieve this, we solve Eq. (\ref{eq73}) for $\cos \beta$, and obtain
\bw
\be\label{eq106}
\cos \beta = \frac{{p_\alpha} {p_\gamma} - \tau \sqrt{p_\beta^2 [p_\alpha^2 -p_\gamma^2+p_\beta^2 \tan ^2(\alpha -\Psi ) ]\tan ^2(\alpha -\Psi )}}{p_\alpha^2+p_\beta^2 \tan ^2(\alpha -\Psi )}, 
\ee
\ew
where $\tau $ is given by Eq. (\ref{eq116}). We substitute Eq. (\ref{eq106}) in Eq. (\ref{eq42}), and obtain $\Phi =p_\alpha^2+p_\beta^2/\cos ^2(\alpha -\Psi )$, which yields
\be\label{eq107}
p_\beta = \varsigma \cos(\alpha - \Psi) \sqrt{\Phi - p_\alpha^2}.
\ee
 We substitute Eq. (\ref{eq107}) in (\ref{eq106}) and obtain (\ref{eq110}), where $\Lambda$ is given by Eq. (\ref{eq115}). Finally, we substitute Eqs. (\ref{eq107}) and (\ref{eq110}) in Eqs. (\ref{eq105}) and (\ref{eq111}), and obtain Eqs. (\ref{eq113}) and (\ref{eq114}). 

}

\acknowledgments

We would like to thank A. Barra, E. Caglioti, U. Locatelli, G. Pinzari and M. Testa for useful discussions.

\bibliographystyle{unsrt}

\end{document}